# Random Laser Emission at Dual Wavelengths in a Donor-Acceptor Dye Mixture Solution

SUNITA KEDIA,* SUCHARITA SINHA

Laser and Plasma Technology Division, Bhabha Atomic Research Centre, Mumbai 400 085, India

*Corresponding author: skedia@barc.gov.in

The work was aimed to generate random laser emissions simultaneously at two wavelengths in a weakly scattering system containing mixture of binary dyes, rhodamine-B (Rh-B) and oxazine-170 (O-170) dispersed with ZnO nano-particles served as scattering centres. Random lasing performances for individual Rh-B dye were extensively studied for varying small signal gain/scatterer density and found lasing threshold significantly depend upon number density of dispersed nano-particles. In spite of inefficient pumping, we demonstrated possibility of random lasing in O-170 dye solution on account of resonance energy transfer from Rh-B dye served as donor. At optimum concentrations of fluorophores and scatterer in dye mixture solution, incoherent random lasing was effectively attained simultaneously at two wavelengths centered 90 nm apart. Dual-emission intensities, lasing thresholds and rate of amplifications were found to be equivalent for both donor and acceptor in dye mixture solution.

## 1. Introduction

Absorption of stimulated emission and scattering of photons are known to decrease efficiency in a conventional laser system. However, in a disordered medium having gain, multiple scattering works positively to increase both oscillations and amplification. A gain medium dispersed with nano-scatters can correlate multiple scattering of photons and can provide feedback within the active medium, thus resulting in light amplification. In such a medium at threshold when gain exceeds loss, non-directional lasing takes place leading to random lasing (RL) [1]. Depending upon the nature of feedback mechanism RL can be either coherent [2] or incoherent [3]. In case of coherent or resonant feedback, the mean free path of light is close to the emission wavelength, and results in localization of radiation field in the medium. This is characterized by a strongly scattering regime, and above threshold the emission intensity increases dramatically with discrete laser modes appearing in the emission spectrum [4-5]. In case of incoherent or non-resonant feedback, light scattering mean free path is much longer than the emission wavelength. This belongs to a weakly scattering regime and results in an increased photon lifetime within the active medium. For non-resonant feedback above threshold, the emission spectrum narrows down to a single peak overriding a broad emission background [6-7]. In case of an active RL system, amplification (gain) and scattering occurs from same element [8] however in a passive system scatterers are present as suspension in the gain medium [6-7]. In a passive RL system, gain per unit length and the number of scatterers are separate quantities and can be controlled independently. It is advantageous to work in such a system as it provides additional freedom to control lasing conditions.

RL action was first discussed by Ambartsumyan *et al* in 1966 [9] and demonstrated by Lawandy *et al* in 1994 [10], and since then this phenomenon has gained revolutionary interest in science because of its simple preparation method, compact size and mirror free cavity configuration. Along with promising ideal illumination source other applications of RL are micro-lasers, speckle-free laser imaging [11], spatial cross talk, time resolve microscopy [12], and cancer detection [13]. In last few years, RL has been demonstrated in rhodamine-6G dye doped polyurethane dispersed with $ZrO_2$ nanoparticles [14], in quantum dots deposited into micro-scale grooves on glass [15], in rhodamine- 640 dye dispersed with silica nanoparticles [7], in ZnO powder produced by sol-gel technique [8], in dye embedded silica gel [16], in single crystalline synthetic opal infiltrated with dye [17], in cholesteric liquid crystal dispersed with silver nanoparticles and laser dye [18] and in rhodamine-B dye dispersed with titanium oxide nano-particles, nano-rods and nano-tubes [6]. Recent developments in this field are designs of multi-coloured RL system [19], tunable RL [20], FRET based RL [21-23] and plasmonic RL system [24]. In most of these reports of RL where two emitters were used as donor-acceptor pair, the lasing was obtained either from donor or from acceptor [20, 21, 24, 25], apart from dual-colour coherent lasing observed in rhodamine-6G and oxazine dye mixture solution dispersed with gold-silver bi-metallic porous nanowires [19].

Here we report our work, where simultaneous dual RL is achieved in a simple passive system containing binary dye mixture solution serving as gain medium dispersed with well studied and easily available zinc oxide (ZnO) nano-scatterers. The chosen dyes were rhodamine-B (Rh-B) and oxazine-170 (O-170). Selection of these two dyes was based on our earlier study which reported the possibility of resonance energy transfer between these two chromophores [26]. Frequency-doubled Nd:YAG laser at 532 nm was employed to pump the gain medium and emissions from the samples were collected using an optical fibre based spectrometer. First set of experiment was focused to attain spectral narrowing from each individual dye solution dispersed with ZnO scatterers. High absorption cross-section associated with pump source in Rh-B ensured easy onset of RL. Lasing performance of Rh-B dye on extent of scattering and gain per unit length in the system were studied by systematically increasing number density of ZnO particles and dye concentration, respectively. A significant dependence of lasing threshold on scatterer density was observed however, lasing threshold remained unchanged when Rh-B dye concentration was varied.

Because of unconventional excitation wavelength, RL could not be obtained in O-170 dye by direct pumping. The inefficient gain at 532 nm for O-170 was overcome by using this dye as an acceptor. An appropriate concentration of Rh-B dye acting as donor was used along with O-170 and spectral narrowing of O-170 emission was reached on account of resonance energy transfer process from Rh-B to O-170. Energy transfer between these emitters was possible because of the overlapping between the emission band of the donor with absorption band of the acceptor. Along with direct pump, the acceptor gained energy from the de-excited photons of Rh-B and could show spectral narrowing. Then we optimized the experimental conditions to reach a situation where both donor and acceptor lased together at two well resolved wavelengths with single excitation at 532 nm. Gain length and scattering density of the system were chosen such that, along with energy transfer to acceptor to reach lasing, the donor could also get feedback and could lase. Two peaks ~92 nm apart and centered at 573 nm and 666 nm associated with Rh-B and O-170, respectively were clearly observed at nearly same lasing threshold. Rate of amplification which was estimated from the slope of linear rise of laser output for high pump energy levels was found to be similar for both dyes [21]. Hence, we demonstrated possibility of controlling random lasing threshold by varying scatterer density. We showed the opportunity of tuning lasing wavelength from 573 nm to 666 nm (about 92 nm shift) in dye mixture solution simply by changing the donor concentration. In addition, we also shown the prospect of multi-colour random lasing in a simpler way in comparison to complicated systems used in earlier reports.

## 2. Experiment

Rh-B and O-170 dyes were separately dissolved in ethanol at milli-molar concentration. Known weight of ZnO particles of size ~50 nm were ultrasonically dispersed in ethanol by continuously stirring the solution for 30 min. Optimized concentration of dye solution dispersed with required concentration of scatters was taken in a 10 mm cuvette and pumped with second harmonic Q-switched Nd:YAG laser output at wavelength of 532 nm (6 ns, 10 Hz). The experimental arrangement for the RL characterization is shown in Fig. 1. The pump beam was focused on the sample using a 50 cm focal length lens (L). Pump laser power incident on the sample was controlled using appropriate neutral density filter (NDF). A 532 nm notch filter was kept in front of optical fibre. Care was taken to avoid feedback from the cuvette windows. Spectral analysis of emitted radiation was performed by collecting emissions at an angle of $30^o$ with respect to incident pump beam, using an optical fibre based spectrometer (Avantes, AvaSpec-2048XL). Cuvette and detector positions were kept undisturbed during the experiment. Nano-particle dispersed dye samples were thoroughly stirred before collection of each spectrum.

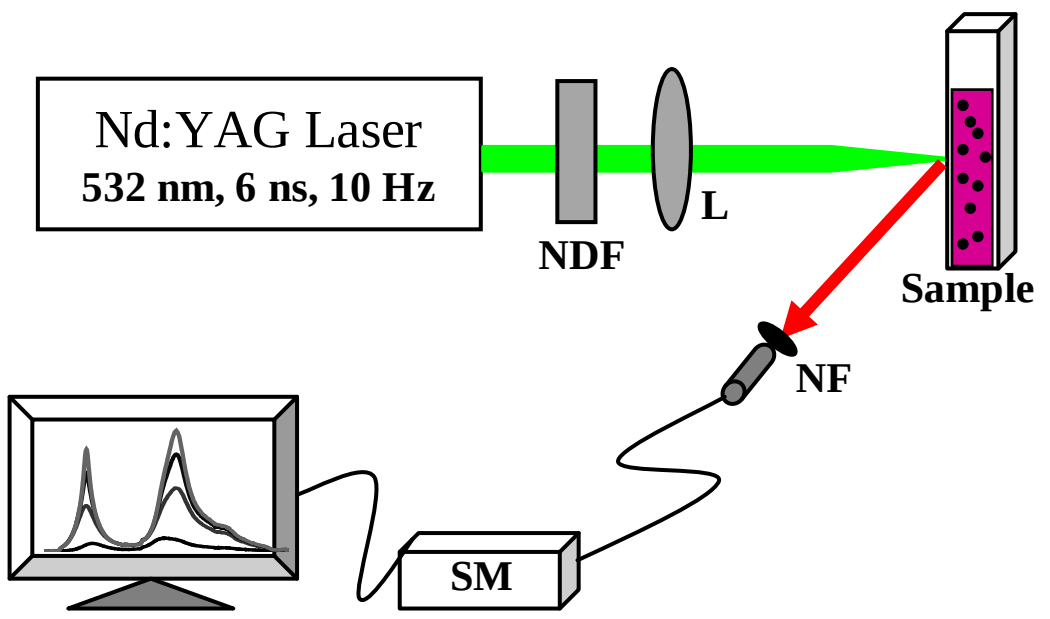

*Fig.1. Experimental arrangement to characterize RL in dye solution immersed with ZnO nanoparticles, where, NDF is neutral density filter, L is 50 cm focal length lens, NF is 532 nm notch filter and SM is spectrometer.*

The RL performance was investigated for different concentrations of dyes and ZnO particles. Series of emission spectra for each combination were recorded for increasing levels of pumping. RL characterization was done by measuring lasing threshold, peak emission wavelength and spectral line width of emission spectra as a function of pump energies.

## 3. Results and Discussion

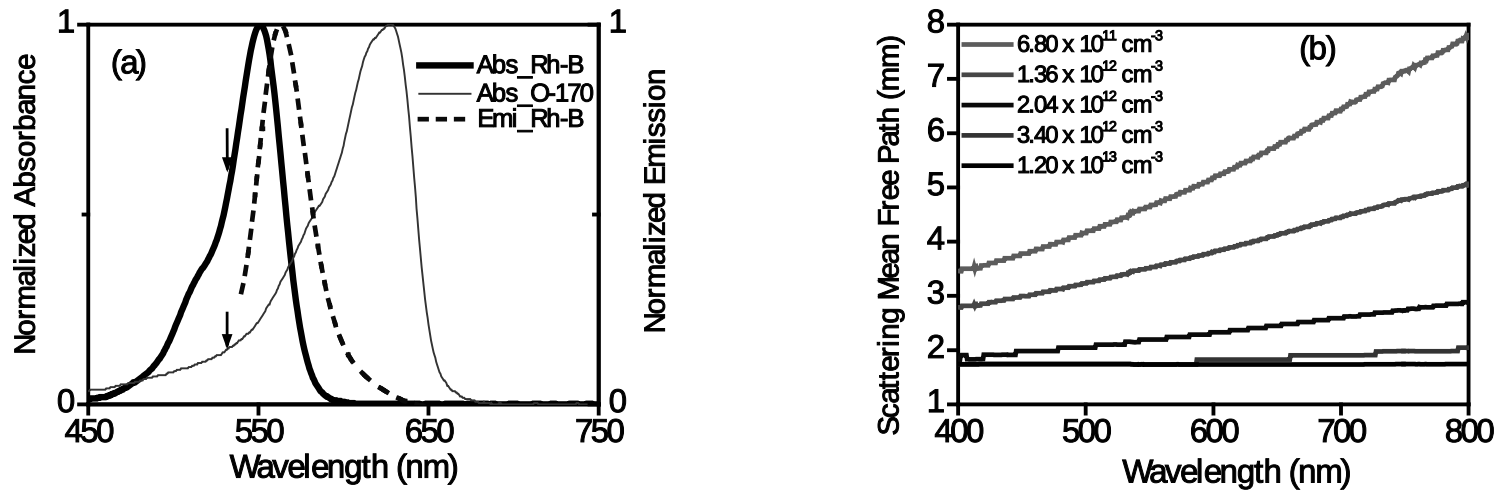

*Fig. 2. (a) Normalized absorbance of Rh-B (thick line) and O-170 (thin line) dye, down arrows marked at 532 nm, normalized emission spectrum of Rh-B dye (dotted line)and (b)scattering mean path for varying concentration of ZnO particles as a function of wavelength*

Thick and thin lines in Fig. 2 are the normalized absorption spectra of Rh-B and O-170 dyes in ethanol solution, respectively. The spectra centered at 551nm and 621 nm correspond to absorption maxima of Rh-B and O-170, respectively. While Rh-B showed significant absorption at 532 nm (marked with arrows), absorption at 532 nm is almost 4 times weaker in case of O-170 dye solution. Dotted line in the figure depicts emission spectrum of Rh-B with emission maximum at 563 nm. There exists a reasonable overlap between emission band of Rh-B and absorption band of O-170. This satisfies an important requirement for energy transfer to take place between these two dyes [26]. To reach RL condition, the dye solution was immersed with varying density of ZnO nanoparticles. Change in scattering mean free path of the system for different concentration of ZnO was estimated using following approach [27]

$$\textit{Mean Free Path} = \frac{T}{\ln(I_1/I_2)} \qquad (1)$$

Here, T is the thickness of the sample which is 10 mm (cuvette thickness), $I_1$ transmitted intensity through pure solvent (ethanol) without gain or scatterers and $I_2$ is transmitted intensity of solvent suspended with known concentration of ZnO particles. Here, $I_1$ and $I_2$ were measured in spectrophotometer using a white light source. Wavelength dependent mean free paths for different concentration of ZnO particles are shown in Fig. 2b. The values were in the range of few millimetres which reduced as concentration of ZnO nanoparticles increased. For coherent feedback, the mean free path typically remains comparable to the emission wavelength [2], characterizing a strongly scattering condition. However, Chen et al reported coherent lasing in dye solution dispersed with ZnO nanoparticles with mean free path compared with our present case [27]. Otherwise in most of the reported work, incoherent random lasing has been observed for weakly scattering regime [6, 7]. In our present case photon mean free path greatly exceeds emission wavelengths of used dyes. Hence, this RL system is expected to behave as a weakly scattering medium. Given the fact that our system is a weakly scattering medium random lasing is expected on account of incoherent feedback rather than localization of radiation field and associated coherent interference effects.

### *A. Dependence on particle density*

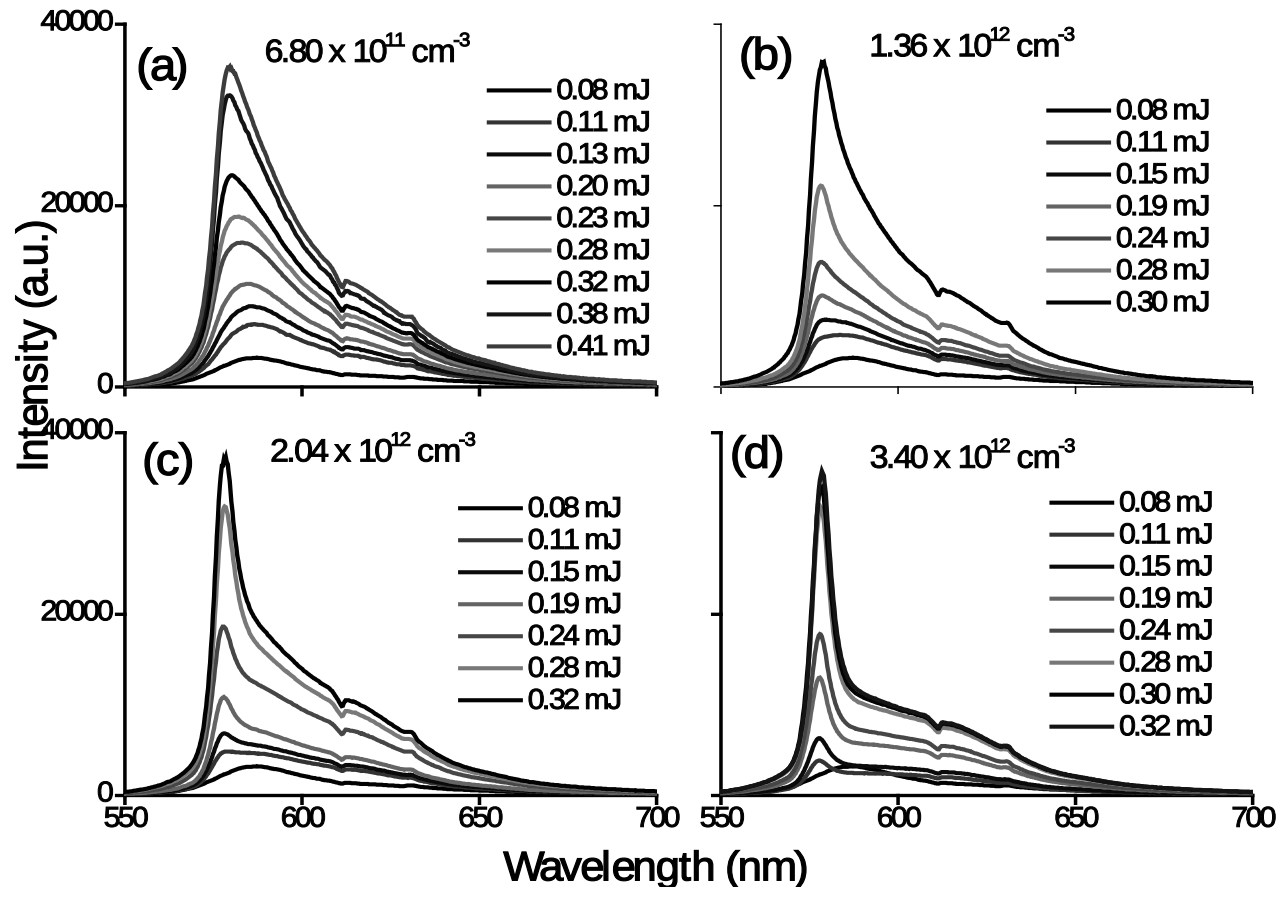

*Fig. 3. Emission spectra of Rh-B dye solution at concentration 1.65 x $10^{-3}$ M dispersed with (a) 6.80 x $10^{11}$ $cm^{-3}$, (b) 1.36 x $10^{12}$ $cm^{-3}$ (c) 2.04 x $10^{12}$ $cm^{-3}$ and (d) 3.40 x $10^{12}$ $cm^{-3}$ number of ZnO particles*

In Figs. 3a-3d, are the emission spectra recorded with Rh-B dye at concentration of 1.65 x $10^{-3}$ M dispersed with different concentrations of ZnO nanoparticles. For low pump energy typically in the region of 0.08 mJ, broad emission of the dye with peak ($\lambda_{max}$) at 587 nm and full width at half maximum (FWHM = $\Delta\lambda$) about 35 nm was observed. This is the spontaneous emission of Rh-B dye. The shift in peak emission wavelength from 563 nm (thin solid line in Fig. 2a) to 587 nm (black curve in Fig. 3a-3d) is because of higher concentration and hence re-absorption of Rh-B dye in the latter case. For lower scatterer concentrations, such as 6.80 x $10^{11}$ $cm^{-3}$ in Fig. 3a, emission intensity increased as the pump level was raised. However, RL was not observed. As nanoparticle concentration was increased, a single narrow peak appeared on the top of the broad spontaneous emission background, as can be seen in Fig. 3b and Fig 3c. The spectral narrowing observed in Fig. 3b indicated the onset of stimulated emission which was further enhanced in Fig. 3c. Number density of the scatter in gain is an important factor as it decides number of scattering events and the mean free path as well as the residence time of the photons in the medium. For low concentration of scatters (Fig. 3a), majority of the photons propagated through the gain medium without experiencing any scattering and therefore they could not contribute to the process of stimulated emission from the dye. Hence, only spontaneous emission got amplified with increasing pump as evident from the broad band emissions shown in Fig. 3a. For higher concentration of ZnO (Fig. 3b-3c) the mean free path of the photons decreased and a major fraction of photons faced multiple scattering resulting in feedback, amplification and spectral narrowing, as observed. However, presence of sizeable spontaneous emission background in these cases indicate, insufficient feedback and amplification not strong enough to extract all optical gain in the form of stimulated emission. With further increase in scatterer density to 3.40 x $10^{12}$ $cm^{-3}$ spectral narrowing of emission spectra sets in at relatively lower pump energy levels. Enhanced stimulated emission is associated with inhibited spontaneous emission, as evident in Fig. 3d. For this concentration of scatterers, the mean free path for photons reduced such that most of the photons experienced multiple scattering and participated in amplification [28]. Associated extended residence time of photons within the medium resulted in improved feedback and therefore relatively diminished spontaneous emission background was observed. Lasing at single wavelength 578 nm with broad emission background indicated incoherent feedback occurred in this system.

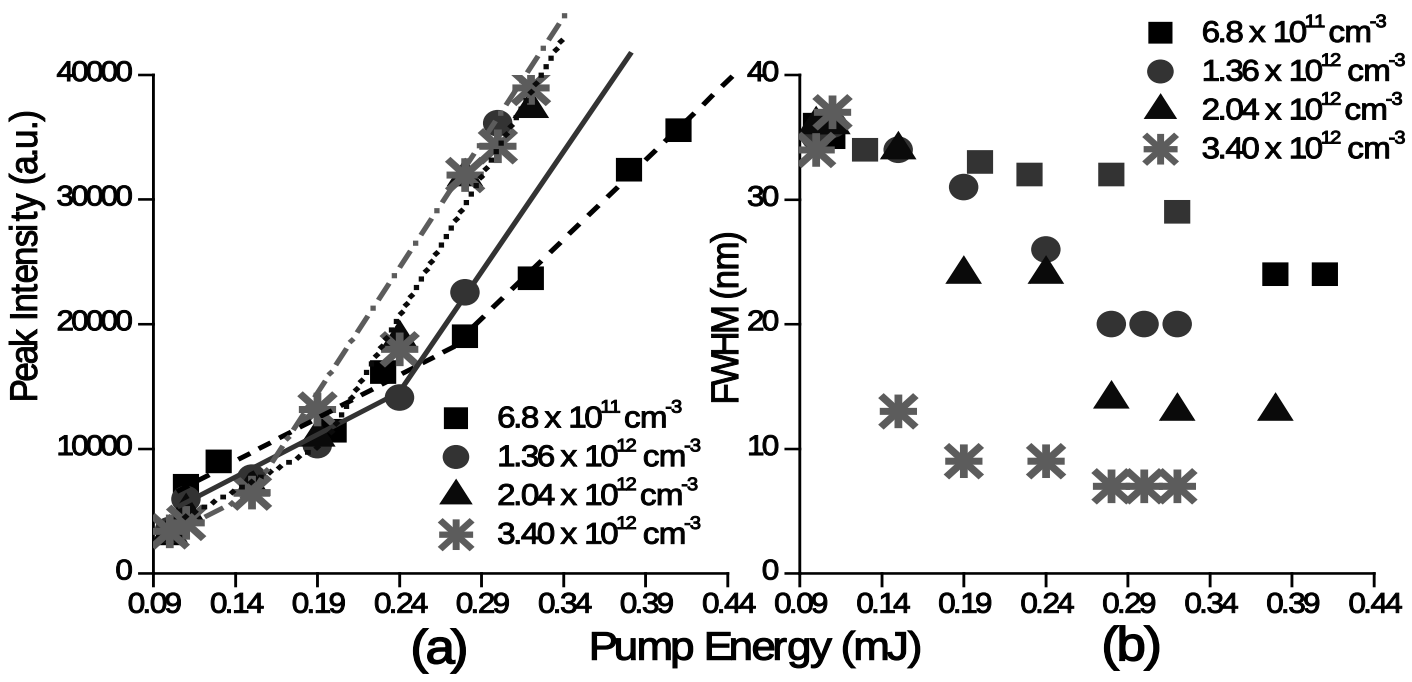

*Fig. 4. (a) Peak emission intensities, and (b) FWHM of emission spectra of Rh-B dye solution at concentration 1.65 x $10^{-3}$ M dispersed with different densities of ZnO particles as a function of pump energies*

Fig. 4a shows the variation in peak intensities of the emission spectra of Rh-B solution containing dispersed ZnO nanoparticles as a function of pump energies. Required minimum pump laser energy corresponding to threshold for onset of RL action decreased from 0.28 mJ per pulse to 0.15 mJ per pulse when nanoparticles concentration increased from 6.8 x $10^{11}$ $cm^{-3}$ to 3.4 x $10^{12}$ $cm^{-3}$. With increasing number of scatterers multiple scattering and hence confinement of photons within active medium and effective amplification increased and this resulted in decreased threshold. Variation in the slope of high pump energy emission was observed in Fig. 4a for different concentration of scatterers. From variation in the high energy slopes it was estimated that, as ZnO concentration increased from 6.8 x $10^{11}$ $cm^{-3}$ to 3.4 x $10^{12}$ $cm^{-3}$ the amplification rate increased by doubled [21]. This can be explained on the basis of dependence of laser power output as a function of pumping rate [29]

$$\boldsymbol{P_e = P_s\left(\frac{R}{R_t} - 1\right)} \qquad (2)$$

here, $P_e$ is the total power generated by stimulated emission, $P_s$ is the power going into spontaneous emission at threshold, R is the pumping rate and $R_t$ is threshold pumping rate. The lasing threshold decreased for higher concentration of scatters and therefore the slope which from equation 2 is inversely proportional to required threshold pumping rate is expected to increase as seen in Fig. 4a. Decrease in FWHM of the emission bands with respect to pump energies can be seen in Fig. 4b. For a concentration of 3.40 x $10^{12}$ $cm^{-3}$ of ZnO scatterer particles in Rh-B solution (stars in Fig. 4b) the spectral width was about 35 nm at lower pump energies. However, with the onset of stimulated emission a sharp reduction in $\Delta\lambda$ to 6 nm was observed after which it remained nearly constant for larger pumping levels.

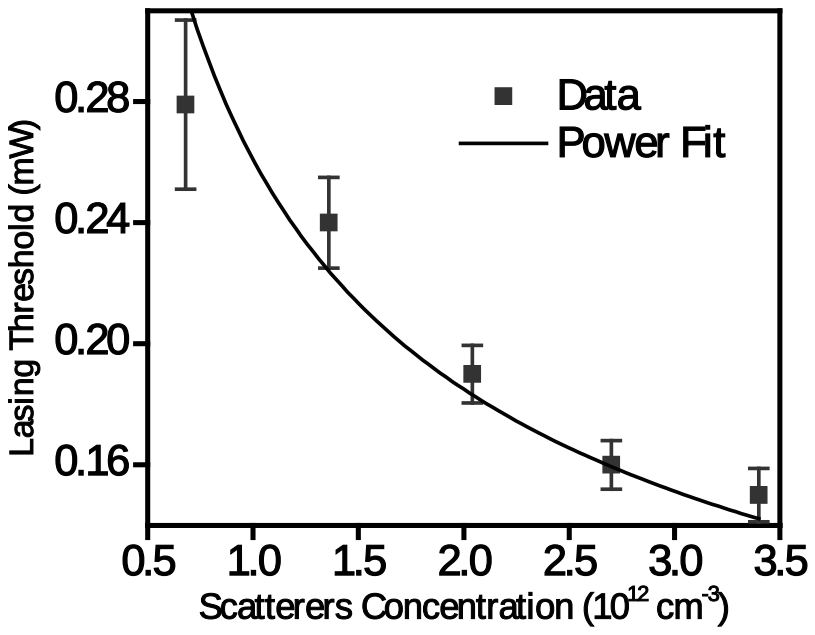

*Fig 5: Random lasing threshold of ZnO dispersed Rh-B dye solution as a function of nanoparticles concentration*

As observed in Fig. 4a, the RL threshold for chosen concentration of Rh-B dye solution decreased when concentration of dispersed ZnO nanoparticles increased. Fig. 5 shows the variation of lasing threshold of Rh-B dye solution as a function of nanoparticle concentration. The threshold depends on number of particle via power function of the form [11]

$$I_{th} = AC^p \qquad (3)$$

Where A is the amplitude factor, C is concentration of ZnO particles and p is the power. The value of p was found as -0.49 by fitting Eq. (3) to the experimental data shown in Fig. 5. This value of power is quite comparable with the value (-0.51) reported by Burin *et al* in their numerical study with light diffusion approximation for a planar two-dimensional system of ZnO disk shaped powder sample [30]. In another numerical investigation by Pinheiro *et al* where lasing threshold of three-dimensional diffusive random laser system composed of point like scatterers was studied, the value of p was obtained as -0.66 [31]. This clearly indicates the dispersed ZnO particles in dye solution behaved more like a two dimensional scattering system in present case.

## B. *Dependence on dye concentration*

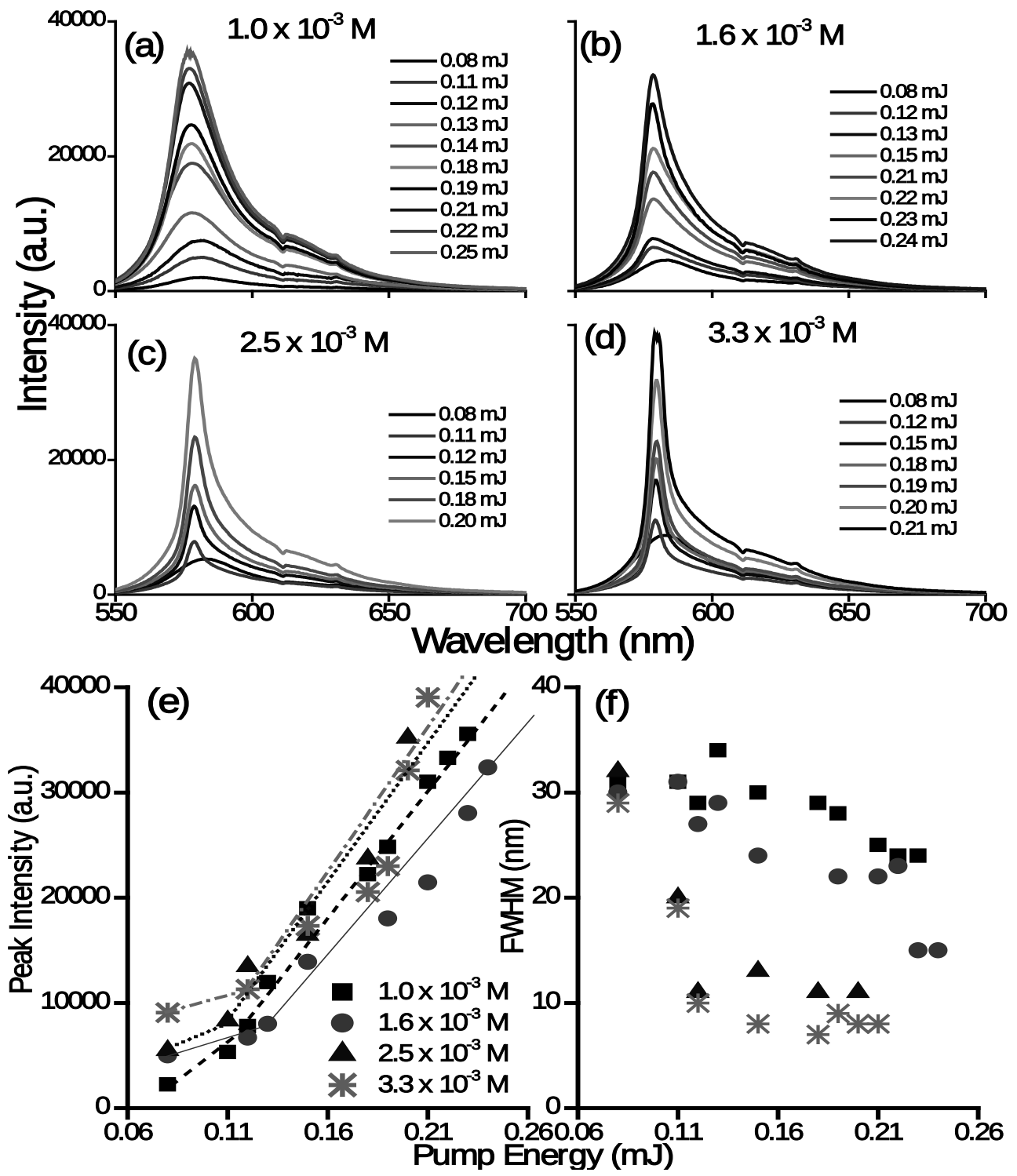

*Fig. 6: Emission spectra of Rh-B dye at concentration (a) 1.0 x $10^{-3}$ M , (b) 1.6 x $10^{-3}$ M ,(c) 2.5 x $10^{-3}$ M ,(d) 3.3 x $10^{-3}$ M dispersed with 1.36 x $10^{12}$ $cm^{-3}$number of ZnO particles, (e) Peak emission intensities of emission bands and (f) FWHM of emission spectra as a function of pump energies*

To explore effect of change in dye concentration on RL, set of emissions from Rh-B dye solutions at different concentrations dispersed with a constant density of ZnO scatterers ~ 1.36 x $10^{12}$ $cm^{-3}$ were recorded, results are shown in Fig. 6a-6d. For lower concentration of dye (Fig. 6a and 6b) RL was not clearly evident although emission did exhibit spectral narrowing around a peak wavelength of 578 nm on addition of scatters in the dye solution. However, distinct onset of build up of stimulated emission along with suppressed spontaneous emission was observed when dye concentration was increased to 2.5 x $10^{-3}$ M and 3.3 x $10^{-3}$ M as shown in Fig. 6c and 6d, respectively. Since, number of scatterers was constant in all these cases, multiple scattering or mean free path of photons was roughly same for all investigated concentrations of Rh-B, Fig. 6a-6d. However, with increasing dye concentration gain per unit length in the system increased and that was primarily responsible for ensuring RL condition to be achieved with threshold being crossed. Rapid build up of stimulated emission is accompanied with suppression of spontaneous emission and gain narrowing facilitated by homogenous nature of broadening of the fluorescent dye based active medium. Change in emission intensities and FWHM with respect to pump energies with varying dye concentration are shown Fig. 6e and 6f, respectively. With constant number of scatterers loss in the system was capped and hence lasing threshold was observed to be nearly same for different dye concentrations (Fig. 6e). For 3.3 x $10^{-3}$ M concentration of dye FWHM of emission spectrum reduced from 31 nm to 7 nm as shown with stars in Fig. 6f.

### *C. RL through energy transfer*

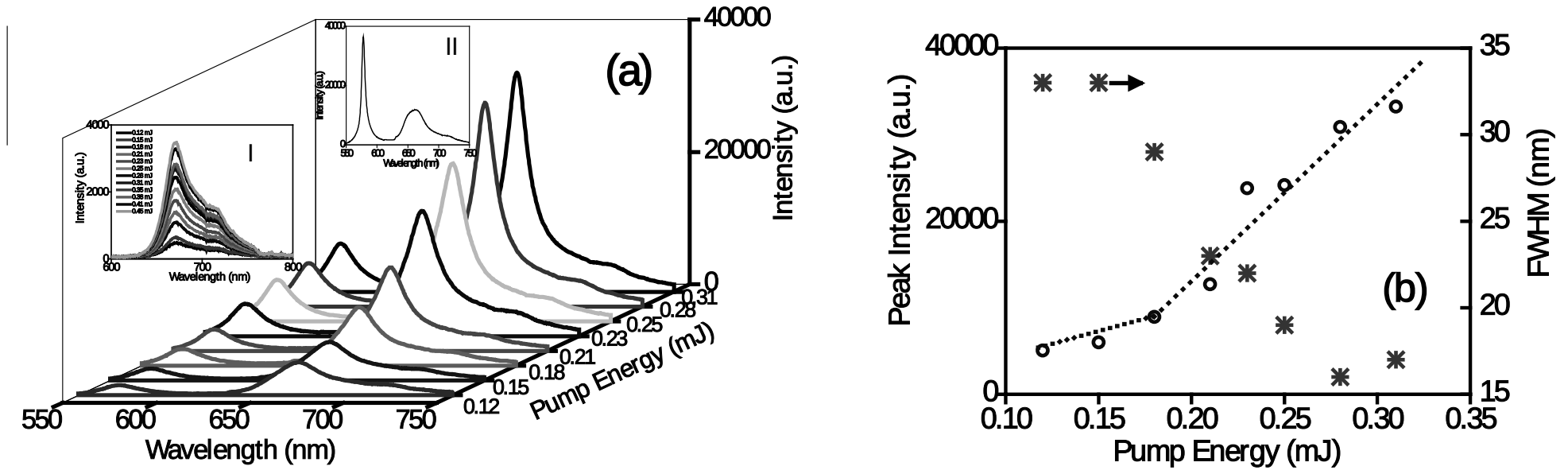

*Fig. 7 (a) Emission of Rh-B and O-170 dye mixture solution with dye concentrations 1.65 x $10^{-3}$ M and 4.5 x $10^{-3}$ M respectively, dispersed with ZnO of 1.36 x $10^{12}$ $cm^{-3}$, inset-I: Emission spectra of O-170 at concentration 4.5 x $10^{-3}$ M dispersed with 1.36 x $10^{12}$ $cm^{-3}$ numbers of ZnO particles, inset-II: Emission of Rh-B and O-170 dye mixture solution with dye concentrations 3.3 x $10^{-3}$M and 4.5 x $10^{-3}$ M respectively ,and (b) Peak intensity and FWHM of O-170 dye in presence of Rh-B at different pump energies.*

As reported elsewhere in Ref-26, poor absorbance of O-170 dye at 532 nm (thin line in Fig. 2a) can be overcome by using this dye as acceptor along with an efficient donor. The donor should have significant absorption at 532 nm and emission band of the donor should overlap with absorption band of the acceptor. Rh-B dye satisfied both these requirements and therefore Rh-B dye at optimized concentrations of 1.65 x $10^{-3}$ M was mixed with O-170 dye at concentration 4.5 x $10^{-3}$ M to make donor-acceptor pair. This dual dye mixture solution was dispersed with 1.36 x $10^{12}$ $cm^{-3}$ concentration of ZnO particles. For low pump pulse energy typically of 0.12 mJ, two peaks centered at 573 nm and 666 nm corresponding to spontaneous emissions from Rh-B and O-170 respectively were observed, can be seen in Fig. 7a. With increased pumping, while emission intensity of both dyes increased, peak intensity of the acceptor was found to enhance much faster than the donor. This occurred since in addition to direct pumping acceptor molecules received energy from photons emitted by de-excited donor. The inefficient gain length of O-170 dye at 532 nm did not allowed this dye to laser even at high dye concentration and comparable high pump energy. The extent of spectral narrowing observed could not observe for an ethanolic solution of individual O-170 dye (inset-II of Fig. 7a) for same concentrations of dye and particles.

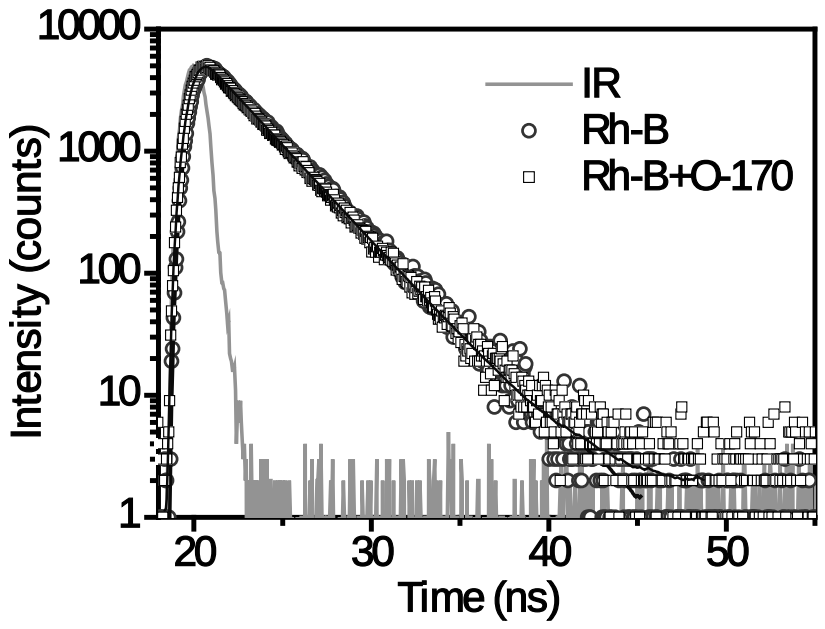

***Fig. 8:*** *Fluorescence decay curve for Rh-B dye solution (circles) and dye mixture solution (squares). 'IR' is the instrument response function.*

In general, the energy transfer from a donor to an acceptor can occur through radiative or non-radiative energy transfer mechanism [32]. In the present case, most of the energy associated with Rh-B is radiatively transferred to O-170 and finally contributes to lasing emission at 666 nm. The fluorescence decay curve of Rh-B dye solution in absence (circles) and presence (squares) of acceptor are shown in Fig. 8. The lifetime of donor changed from 2.85 ns to 2.75 ns in presence of the acceptor. This insignificant reduction in fluorescence lifetime of donor indicates the energy transfer between these chromophores is mainly by radiative pathway. The extent of quenching in the spectral overlapping part of Rh-B dye in presence of acceptor supported this fact [33]. The efficient energy transfer between these dyes occurred only at optimized concentration of donor. At other higher concentration of Rh-B donor such as 3.3 x $10^{-3}$ M in the donor-acceptor mixture emission from Rh-B dye dominated, as can be seen in the inset-II of Fig. 7a. Hence, in this donor-acceptor pair the lasing mode could be controlled by varying the concentration of donor. The lasing wavelength moved from 573 nm for Rh-B to 665 nm for O-170 dye, about 92 nm shift when donor concentration was simply reduced from 3.3 x $10^{-3}$ M (inset-II) to 1.65 x $10^{-3}$ M in Fig. 7a. In presence of the donor, spectral width of acceptor emission decreased from 33 nm to 15 nm (stars in Fig. 7b).

*D. Dual emission in dye mixture solution*

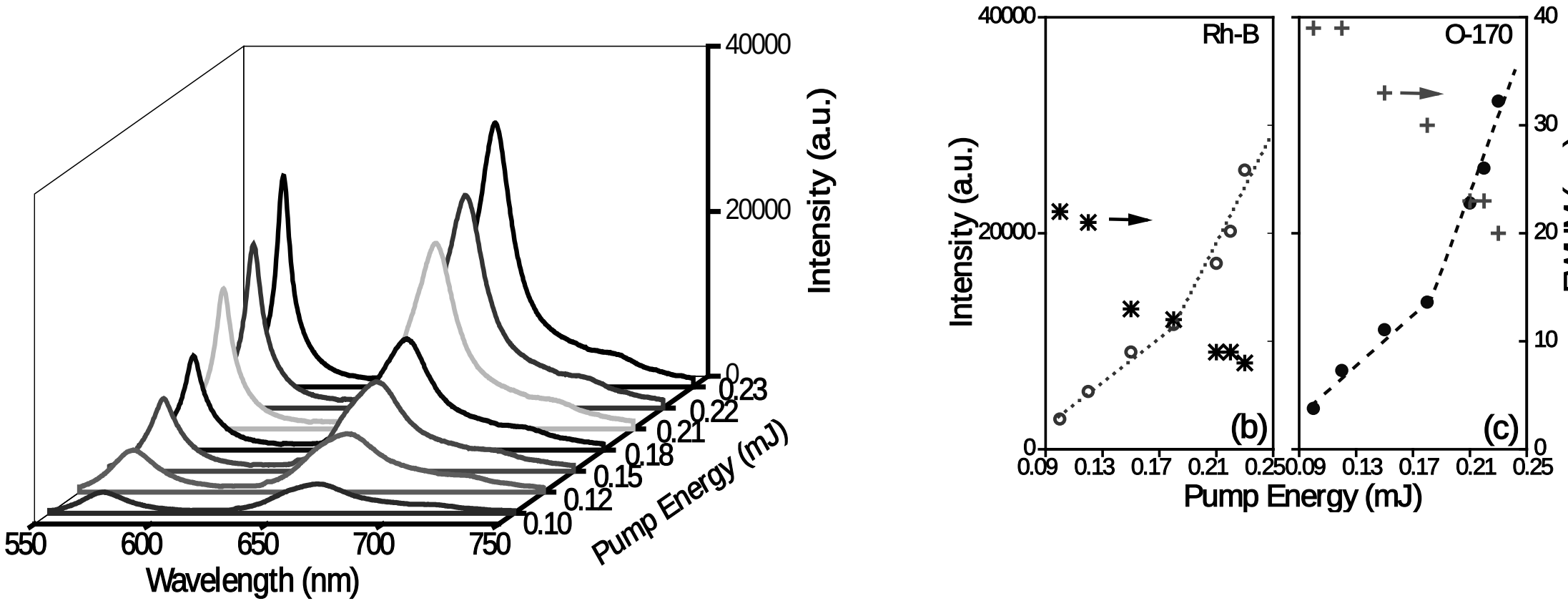

*Fig. 9: (a) Emission spectra of Rh-B and O-170 dye mixture solution at concentration 4.20 x $10^{-3}$ M and 9.0 x $10^{-3}$ M, respectively immersed with 1.22 x $10^{13}$ $cm^{-3}$ ZnO nanoparticles and variation of FWHM and peak intensities of the spectra with respect to pump energies of (b) Rh-B and (c) O-170*

The concentrations of Rh-B and O-170 dyes were systematically optimized to 4.20 x $10^{-3}$ M and 9.0 x $10^{-3}$ M, respectively, such that effective gain for emission from both dyes was similar. The mixture was dispersed with 1.22 x $10^{13}$ $cm^{-3}$ ZnO nanoparticles and emissions were collected at different pump levels. Two intense spectral bands associated with Rh-B and O-170 were clearly observed which narrowed down at high pump energy, as shown in Fig. 9a. As there was an incomplete overlapping between the emission band of Rh-B and absorption band of O-170, only a fraction of energy from Rh-B got transferred to O-170 and this energy was enough to support random lasing from O-170 acceptor dye. The residual energy in Rh-B got amplified along with feedback from scattering and stimulated emission from Rh-B dye was observed. At threshold around 0.17 mJ, incoherent random lasing emission at dual wavelengths 573 nm and 664 nm were simultaneously observed from Rh-B and O-170, respectively when pumped at 532 nm. FWHM of Rh-B reduced from 21 nm to 8 nm when pump energy was increased from 0.10 mJ to 0.23 mJ, Fig. 9b. However, for the similar change in pump energies spectral width of O-170 reduced from 39 nm to 20 nm, as shown in Fig. 9c. Simultaneous presence of absorbing acceptor dye O-170 reduced the effective gain over emission spectra of Rh-B. Reduced effective gain leads to limited gain narrowing for emission characteristic of dye Rh-B. Hence, spectral narrowing for pure Rh-B dye was evidently more than for donor-acceptor dye mixture Rh-B and O-170, all other factors remaining same. Amplification rate as estimated from the slope of liner rise of laser output for high pump energy levels were found nearly same for both Rh-B and O-170 dyes [21].

## Conclusion

We successfully demonstrated the random lasing emission in Rh-B and O-170 dye solutions using dispersed ZnO nano-particles as scaterers. Decrease in lasing threshold was observed when scatterer density was increased in Rh-B dye solution. In spite of marginal absorption at pump wavelength random lasing was effectively achieved in O-170 dye aided by energy transfer process using Rh-B as donor. Unchanged fluorescence lifetime and quenching in donor fluorescence intensity in the spectral overlap region confirmed the energy transfer between these dyes was through radiative path way. Dyes concentrations were optimized to make small signal gain for emission from both dyes comparable and at appropriate concentration of scatterer, incoherent random lasing was successfully attained simultaneously at two wavelengths arising from the two individual dyes.

**Acknowledgment**. Authors acknowledge Dr. B. Nilotpal, RPCD, BARC for his help in lifetime measurement.